\documentclass[conference]{IEEEtran}
\IEEEoverridecommandlockouts
\usepackage{threeparttable}

\usepackage{cite}
\usepackage{amsmath,amssymb,amsfonts}
\usepackage{algorithm}
\usepackage{algpseudocode}
\usepackage{graphicx}
\usepackage{textcomp}
\usepackage{xcolor}
\def\BibTeX{{\rm B\kern-.05em{\sc i\kern-.025em b}\kern-.08em T\kern-.1667em\lower.7ex\hbox{E}\kern-.125emX}}

\begin{document}
\title{{CIM-Tuner: Balancing the Compute and Storage Capacity of SRAM-CIM Accelerator via Hardware-mapping Co-exploration}\\
}

\author{\IEEEauthorblockN{Jinwu Chen$^{1,2}$, Yuhui Shi$^{1}$, He Wang$^{1}$, Zhe Jiang$^{1, 2}$, Jun Yang$^{1, 2}$, Xin Si$^{1, 2 *}$ and Zhenhua Zhu$^{3 *}$}
\IEEEauthorblockA{$^{1}$School of Integrated Circuits, Southeast University, China $^{2}$ National Center of Technology Innovation for EDA, China \\
$^{3}$ Department of Electronic Engineering, Tsinghua University, China \\
Email: chenjinwu@seu.edu.cn, 220236525@seu.edu.cn, wanghe020608@gmail.com, zhejiang.arch@gmail.com, \\ 
dragon@seu.edu.cn, xinsi@seu.edu.cn, zhuzhenhua@mail.tsinghua.edu.cn \\
$^{*}$ Corresponding authors.}
% \IEEEauthorblockA{$^{2}$\textit{Department Name 2} \\
% \textit{Institution Name 2}\\
% City, Country \\
% email2@example.com}
% \IEEEauthorblockA{$^{3}$\textit{Department Name 3} \\
% \textit{Institution Name 3}\\
% City, Country \\
% email3@example.com}
}

\maketitle

\begin{abstract}
As an emerging type of AI computing accelerator, SRAM Computing-In-Memory (CIM) accelerators feature high energy efficiency and throughput.
However, various CIM designs and under-explored mapping strategies impede the full exploration of compute and storage balancing in SRAM-CIM accelerator, potentially leading to significant performance degradation. 
To address this issue, we propose CIM-Tuner, an automatic tool for hardware balancing and optimal mapping strategy under area constraint via hardware-mapping co-exploration. It ensures universality across various CIM designs through a matrix abstraction of CIM macros and a generalized accelerator template. For efficient mapping with different hardware configurations, it employs fine-grained two-level strategies comprising accelerator-level scheduling and macro-level tiling.
Compared to prior CIM mapping, CIM-Tuner's extended strategy space achieves 1.58$\times$ higher energy efficiency and 2.11$\times$ higher throughput.
Applied to SOTA CIM accelerators with identical area budget, CIM-Tuner also delivers comparable improvements. 
The simulation accuracy is silicon-verified and CIM-Tuner tool is open-sourced at https://github.com/champloo2878/CIM-Tuner.git.

\end{abstract}


\section{Introduction}

SRAM Computing-in-Memory (SRAM-CIM) has garnered significant attention as a highly energy-efficient matrix multiplication engine. 
By enabling data movement with shorter distances and larger bandwidth, alongside customized designs for data precision and sparsity, SRAM-CIM achieves notable improvements in both energy and throughput. 
These advantages are validated at both the macro level~\cite{LCC-6T}--\cite{FPCIM}  and accelerator level~\cite{TranCIM}--\cite{Sticker-IM}, positioning SRAM-CIM as a compelling alternative to digital computing arrays.

\begin{figure}[!t]
\centering
\includegraphics[width=3.2in]{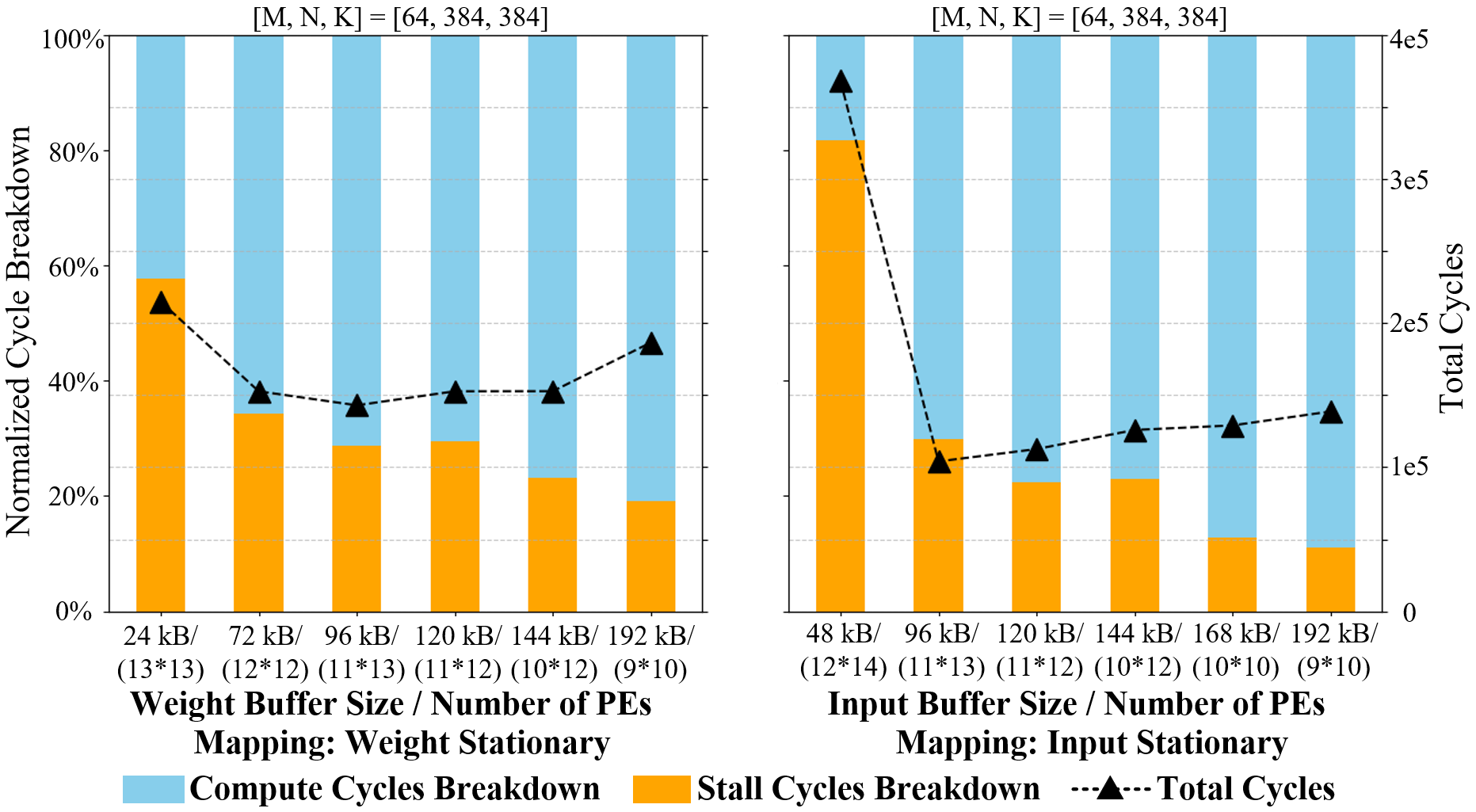}
\caption{Latency breakdown of matrix multiplication on systolic accelerator. Evaluated using scale-sim~\cite{scale-sim}. Left: varying weight buffer in weight stationary mapping. Right: varying input buffer in input stationary mapping. }\label{1M}
\end{figure}

Current SRAM-CIM accelerator optimizations primarily target CIM computation itself, neglecting the compute and storage resources balancing and corresponding mapping strategy matching.
However, this neglect may significantly degrade the inference performance of these DNN accelerators.
Back to the conventional digital accelerators, the size of the on-chip buffer has a significant impact. Figure~\ref{1M} shows the performance of a systolic array under different compute and storage allocations of fixed area budget.
It demonstrates that stall cycles decrease as the buffer size increases, revealing that larger buffer leverages data locality better and improves data reuse to alleviate external bandwidth pressure. 
However, an oversized buffer inevitably reduces the area available for the computation array, leading to lower computation throughput and an overall latency increase.

For SRAM-CIM based accelerator, due to tightly coupled compute and storage units, exploring the compute and storage balancing demands more fine-grained modeling and mapping analysis compared to conventional digital arrays. 
Such exploration requires delving into specific SRAM-CIM macro designs. 
Previous optimization works like AutoDCIM~\cite{Autodcim} focused only on analyzing the storage-compute balance in a specific digital CIM macro template, lacking support for other CIM types and accelerator-level optimizations. 
To address this gap, these two challenges must be tackled:
(1) \textbf{Adaptation to various designs:} Diverse SRAM-CIM implementations vary in data precision, accuracy, and compute mechanisms (e.g., bit-cell design, multi-bit expansion method, simultaneous compute/update capability and so on). This diversity increases the complexity of the problem and may push any solution towards being highly design-specific. 
(2) \textbf{Underexplored mapping strategies:} 
Different storage and compute ratio require distinct optimal mapping strategies for accelerators. 
As a novel architecture, SRAM-CIM accelerator lacks comprehensive exploration of matrix multiplication mapping. 
Especially as current neural network operators vastly exceed on-chip buffer capacity, traditional static mapping proves inefficient, necessitating dynamic and fine-grained strategies to account for frequent parameter updating. 

To address these challenges, we propose CIM-Tuner: an automated and open-sourced tool for balancing compute and storage capacity in SRAM-CIM accelerators. Within given CIM macro configurations and network workloads, CIM-Tuner simultaneously determines the optimal hardware sizing and operators mapping under area budget constraint and optimization target. The contributions include:

\begin{itemize}
\item First, we introduce a matrix abstraction for SRAM-CIM macros and establish a universal accelerator template to decouple diverse circuit implementations from architectural exploration. Unified evaluation formulas adaptable to any SRAM-CIM design are provided.

\item Second, we propose the fine-grained two-level mapping strategies for better hardware resource exploration: \textbf{At the accelerator level,} both spatial and temporal scheduling methods are proposed, encompassing the mapping strategies introduced in previous works. \textbf{At the macro level,} novel Accumulation-First (AF) tiling and Parallel-First (PF) tiling strategies exploiting CIM-specific \textbf{storage-compute ratio (SCR)} to optimize CIM utilization while expanding the mapping strategy design space.

\item Finally, we explore the hardware-mapping co-design space via simulated annealing. To accelerate exploration, the hardware design space is heuristically pruned and the layers with same size are gathered to reduce the mapping choices. 
Experimental results demonstrate that our expanded mapping strategy achieves 1.58$\times$ higher energy efficiency and 2.11$\times$ higher throughput compared to previous CIM mapping methodologies under the same co-exploration process.
Also, evaluated on SOTA CIM accelerators~\cite{TranCIM},~\cite{TP-DCIM} under identical area budgets, CIM-Tuner delivers comparable improvements of 1.34$\sim$2.31$\times$ in energy efficiency and 1.03$\sim$2.88$\times$ in throughput.
\end{itemize}

The simulation accuracy of CIM-Tuner tool is silicon-verified and CIM-Tuner is open-sourced for the purpose of better early stage SRAM-CIM accelerator design decisions.

% mapping relate to SCR

\section{Background and Motivation}

\subsection{SRAM-CIM Basis and Storage-Compute Ratio}
Various CIM technologies leverage different types of memory. Among them, SRAM features high access speed, nearly unlimited endurance and seamless integration with advanced CMOS technology.  Thus, SRAM-CIM has been widely researched and utilized in many DNN accelerators~\cite{TranCIM}--\cite{Sticker-IM}.
% , including commercial applications~\cite{MediaTek}.
Typically, SRAM-CIM operates in two modes: compute and weight update. During compute, input vectors are projected into the CIM to perform multiply-accumulate (MAC) operations with stored weights. The update mode resembles conventional memory write operations.

\begin{figure}[!t]
\centering
\includegraphics[width=3.4in]{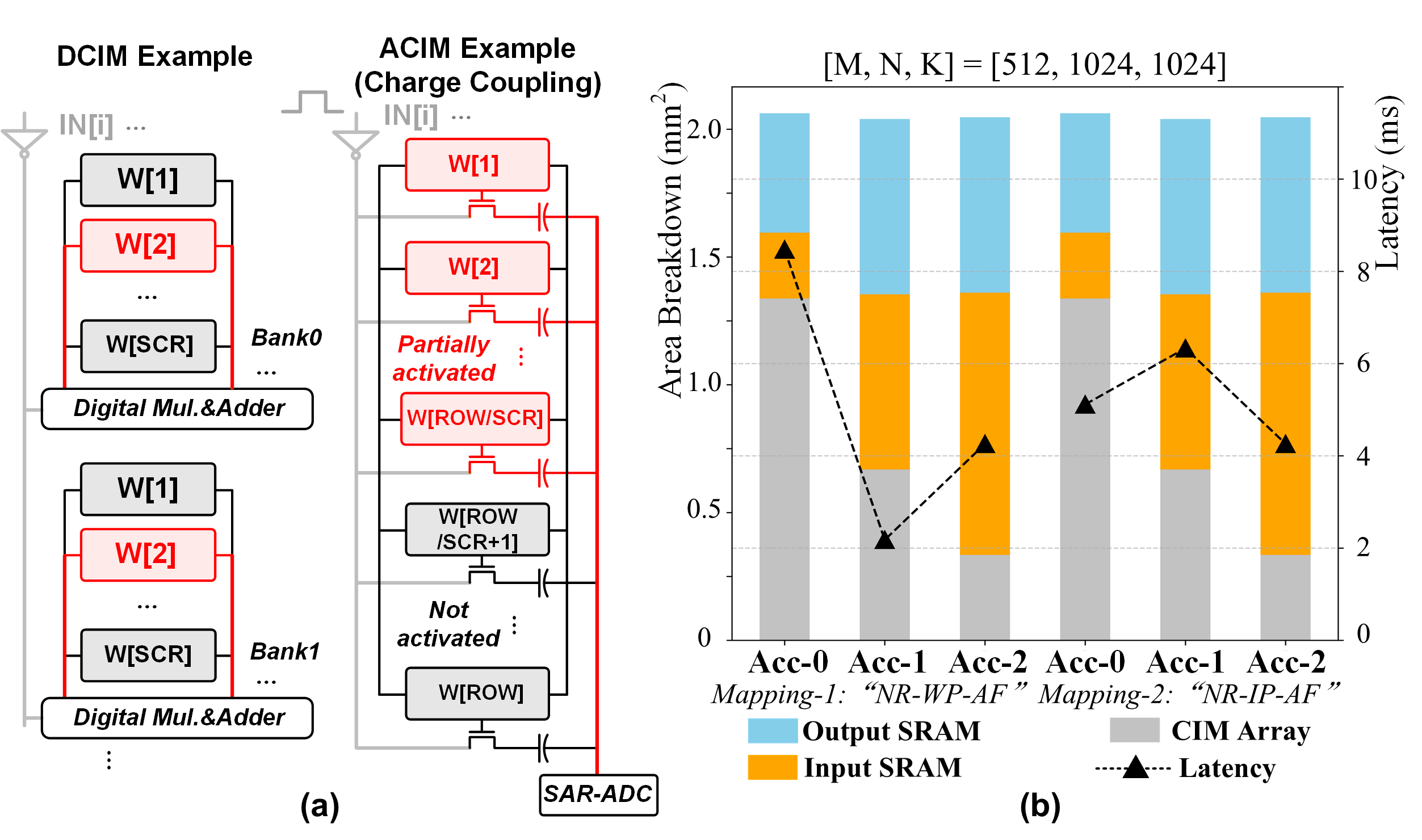}
\caption{(a) The SCR mechanism in both digital and analog CIM. (b) The latency of matrix multiplication across different hardware proportions and mapping strategies. See evaluation setup in Section IV. }\label{2M}
\end{figure}

As shown in Figure~\ref{2M} (a), the storage-compute ratio (\textit{SCR}) in CIM is defined as the ratio of memory cells to compute units within a CIM macro. In digital CIMs (DCIMs), multiple memory cells typically share one set of multipliers and adders within a local bank to maintain storage density and layout compactness~\cite{FPCIM}. Here, \textit{SCR} equals the number of cells per local bank. For analog CIMs (ACIMs), \textit{SCR} represents the ratio of total column cells to activated cells during computation, ensuring sufficient compute-line signal margin~\cite{Sticker-IM}.
In SRAM-CIM accelerators, weight data is typically stored in the CIM array. Input data from different channels reuses weights, generating sums for different output channels. 

\subsection{Motivation for Compute and Storage Balancing}
Currently, with the explosive growth in model parameters, the size of a single matrix multiplication operator ($\sim$MB level) often far exceeds the SRAM capacity of the on-chip SRAM-CIM accelerator ($\sim$KB level)~\cite{111}--\cite{Bert}. This requires frequent updating of input data, weight, and output data during computation. 
Therefore, under limited area budget, selecting the correct ratio of compute and storage capacity, as well as the fine-grained mapping strategies, is particularly crucial.
As shown in Figure~\ref{2M} (b), assigning excessive compute capacity may lead to significant cycles wasted on data movement. Conversely, allocating excessive storage capacity can result in low computation throughput, also causing high latency. More than $4 \times$ worse latency is observed in our motivation demonstration.
Furthermore, different mapping strategies observed under the same hardware configuration may yield vastly different performance. In Figure~\ref{2M}, the input-priority update strategy demonstrates totally different latency variation with the weight-priority update strategy under the same hardware. 
This demonstrates that both the compute and storage balancing and mapping strategies matching can severely impact the performance of CIM accelerators.

However, existing SRAM-CIM accelerator designs~\cite{TranCIM}--\cite{Sticker-IM} mainly focuses on optimizing computation paths leveraging techniques such as data sparsity, mixed-precision computation, and operator fusion. 
Meanwhile, critical decisions concerning the size of on-chip buffers and the storage-compute ratio of CIM macros remain largely reliant on designers' empirical intuition. 
This ad hoc approach often fails to achieve optimal system performance.
Moreover, previous CIM mapping strategy~\cite{SXY-TCAD} is under-explored, leaving the opportunity for higher performance under various hardware configurations.
Therefore, this gap in both hardware balancing and mapping strategy matching motivates us to propose CIM-Tuner, an automated tool for optimal hardware sizing and mapping strategies under given macro configuration and area budget.

\section{CIM-Tuner Framework}

\begin{figure}[!t]
\centering
\includegraphics[width=3.4in]{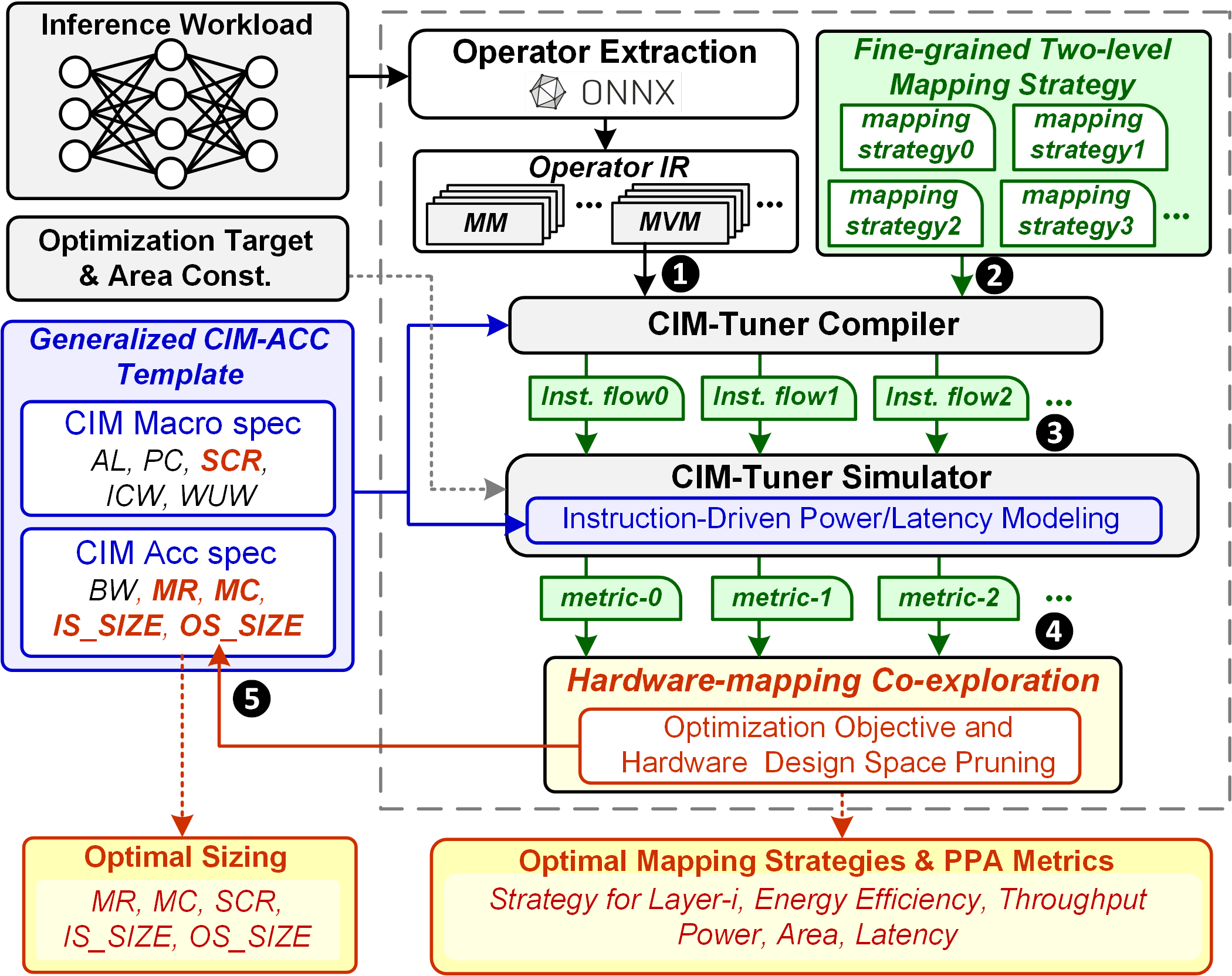}
\caption{Overview of CIM-Tuner's hardware-mapping co-exploration.}\label{Overview}
\end{figure}

\subsection{Overview of CIM-Tuner}
Figure~\ref{Overview} illustrates the workflow of CIM-Tuner, where mapping exploration operates as a sub-process of hardware exploration.
For mapping exploration, first, target workload operators are represented through intermediate representations (IR) to extract matrix dimensions. Second, fine-grained two-level operator mapping strategies are explored. These strategies are converted into unified instruction flows via the CIM-Tuner compiler. Third, using the CIM-Tuner simulator's instruction-driven power and latency models, cycle-accurate performance and power consumption simulations are performed for different mapping strategies. This enables selection of the optimal mapping strategy and generates PPA metrics.

Outside the mapping, the hardware exploration can be easily conducted by modifying the hardware configuration
%  of the generalized template
, with the corresponding notations detailed in Table~\ref{notation}. 
% In this work, Simulated annealing algorithm is used to conduct the pruned co-exploration design space. And finally obtain the optimal hardware configuration paired with corresponding optimal mapping strategy.

\subsection{Matrix Abstraction and Generalized Accelerator Template}

\begin{figure}[!t]
\centering
\includegraphics[width=3.4in]{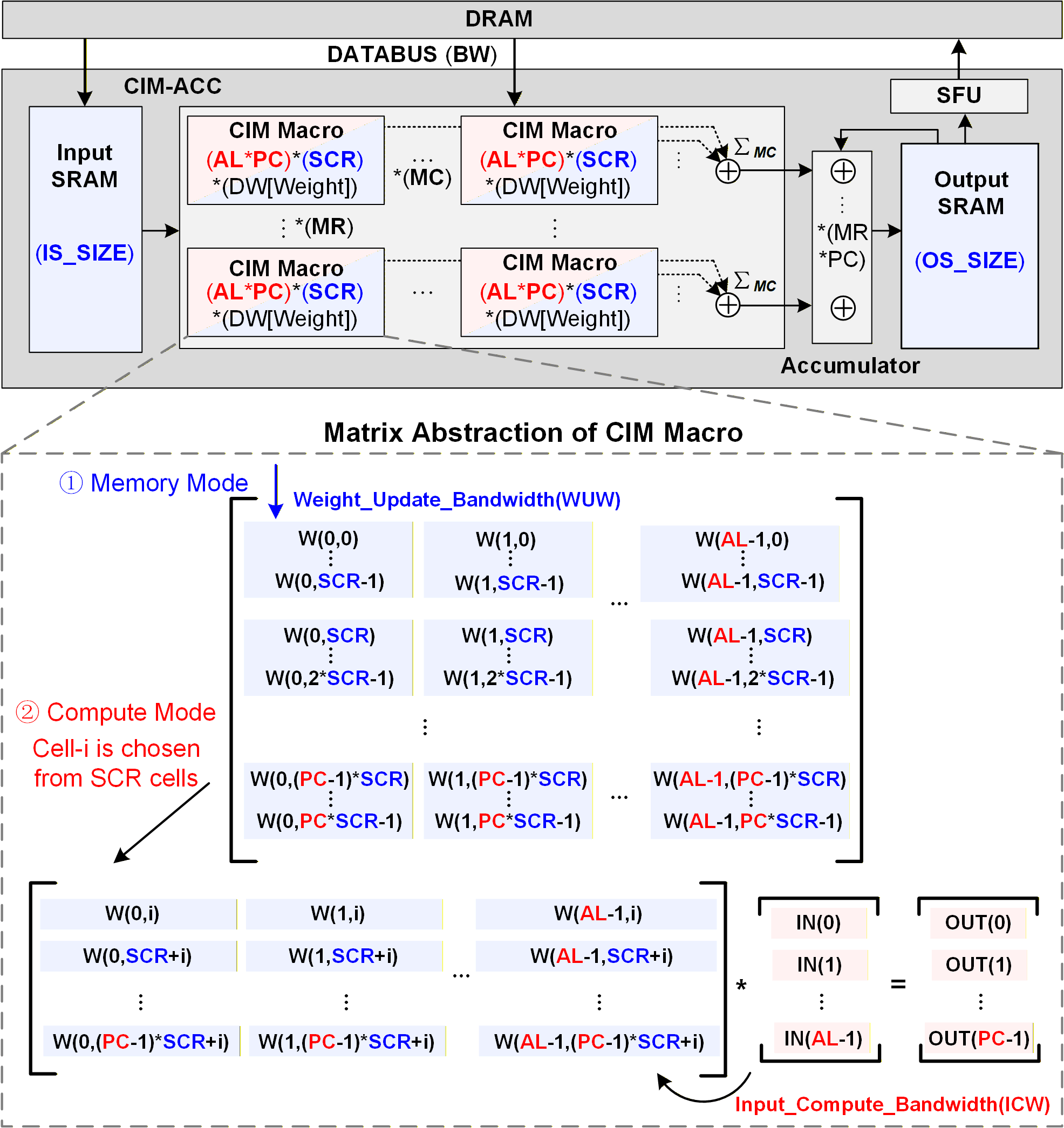}
\caption{Generalized three-stage pipelined CIM accelerator (top) and proposed matrix abstraction of SRAM-CIM macro (bottom).}\label{Template}
\end{figure}

\textbf{Matrix Abstraction of CIM Macro.} 
To circumvent the complex taxonomy of SRAM-CIM micro-architectures, we abstract CIM Macros as a mathematical matrix, as shown in Figure~\ref{Template}. 
We observe that all SRAM-CIM variants fundamentally perform the same atomic operation: a vector-matrix projection between an input vector of accumulation length (\textit{AL}) and a weight matrix of \textit{AL} $\times$ parallel channel (\textit{PC}) stored in SRAM-CIM, generating a partial sum vector of length \textit{PC}. The Storage-Compute Ratio (SCR) mechanism dynamically selects one from \textit{SCR} available $AL\times PC$ weight matrices for computation.
Besides, to standardize computation and weight update latencies across various CIM designs, CIM-Tuner introduces two key parameters: (1) Input computing bandwidth (\textit{ICW}) representing processable input data bits per cycle. (2) Weight update bandwidth (\textit{WUW}) representing updated weight data bits per cycle.

\begin{table}[t!]
\centering
\caption{Notations of Generalized Accelerator Template}
\begin{center}
\begin{tabular}{|c|c|c|c|}
\hline
\textbf{Macro Parameters} & \textbf{Notation} & \textbf{Acc. Parameters} & \textbf{Notation} \\
\hline
Accumulation Length & \textit{AL} & Macro Cols & \textit{MC} \\
\hline
Parallel Channel & \textit{PC} & Macro Rows & \textit{MR} \\
\hline
Storage-Compute Ratio & \textit{SCR} & Bus Bandwidth & \textit{BW} \\
\hline
Input-Compute Bandwidth & \textit{ICW} & Input SRAM & \textit{IS\_SIZE} \\
\hline
Weight-Update Bandwidth & \textit{WUW} & Output SRAM & \textit{OS\_SIZE} \\
\hline
% \multicolumn{6}{l}{$^{\mathrm{*}}$EE. is for optimal energy efficiency and Th. is for optimal throughput. Other hardware parameters are fixed during exploration.}
\end{tabular}\label{notation}
\end{center}
\end{table}

% These two variables provide critical complementary modeling capabilities for diverse CIMs, 
These two variables enable the characterization of data input mechanism across CIM implementations. For example, in DCIM, \textit{ICW} typically correlates with the number of input bitline of one local computing unit, whereas in ACIM, \textit{ICW} is generally associated with the Digital-to-Analog Converter (DAC) precision of one analog input driver:

\begin{equation}
ICW_{DCIM} =  AL \times N_{InputBitline}
\end{equation}
\begin{equation}
ICW_{ACIM} = AL \times Precision_{DAC}
\end{equation}

The latency of single CIM computation is:
\begin{equation}
Latency_{DCIMCompute} = \frac{Datawidth[\textit{Input}]}{ N_{InputBitline}} \times \frac{1}{\textit{freq}}
\end{equation}
\begin{equation}
Latency_{ACIMCompute} = \frac{Datawidth[\textit{Input}]}{ Precision_{DAC}} \times \frac{1}{\textit{freq}}
\end{equation}

The \textit{WUW} is associated with the specific bitcell design and multi-bit extension method. The latency of single CIM update can be characterized by:

\begin{equation}
Latency_{update} = \frac{AL \times Datawidth[\textit{Weight}]}{WUW} \times \frac{1}{\textit{freq}}
\end{equation}

\textit{freq} represents the operating clock frequency. Thus the matrix abstraction bridges hardware design specifics and element-wise arithmetic operations. By incorporating the reported energy efficiency, area, and frequency, various SRAM-CIM designs can be accurately modeled. Compared with previous fixed macro design search~\cite{Autodcim}, matrix abstraction effectively supports complex circuit implementation details.

\begin{figure}[!t]
\centering
\includegraphics[width=3.4in]{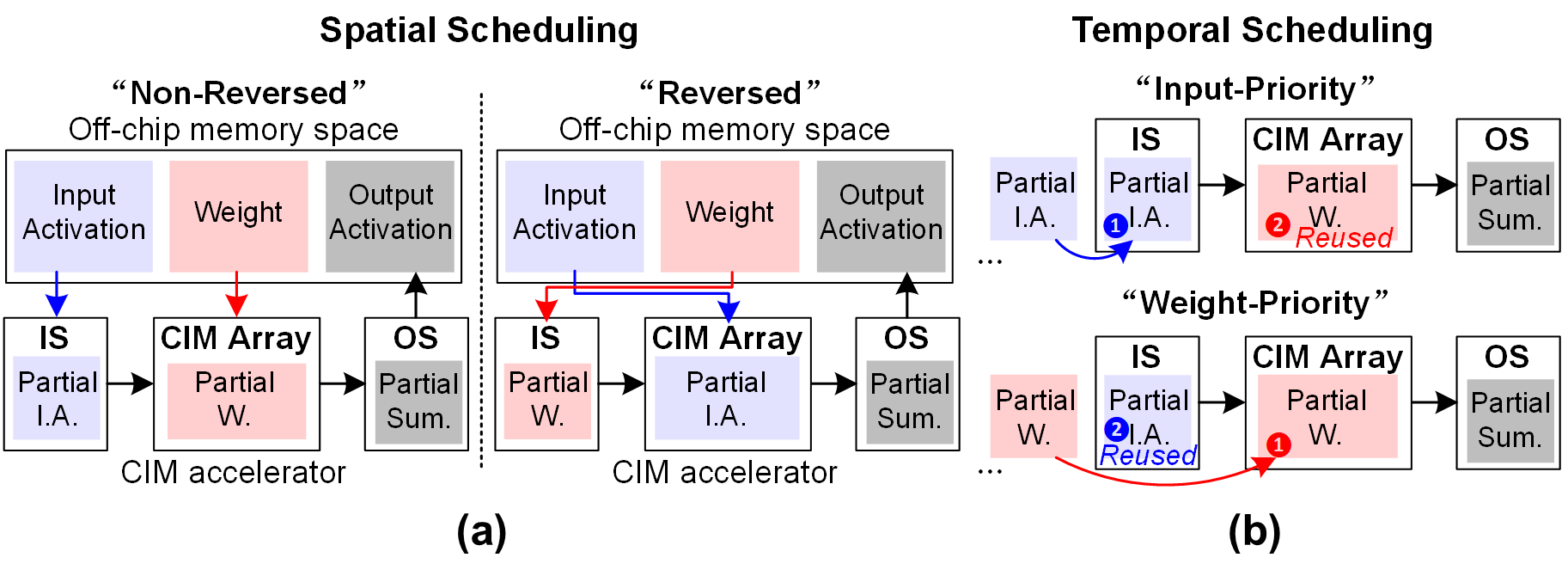}
\caption{Accelerator-level scheduling strategies, IS and OS are short for Input SRAM and Output SRAM, respectively: (a) Non-Reversed (NR) and Reversed (R) spatial scheduling.\ (b) Input-Priority (IP) and Weight-Priority (WP) temporal scheduling.}\label{Scheduling}
\end{figure}

\textbf{Generalized Accelerator Template.} 
Unlike traditional digital computing arrays that utilize scratchpad memory to distribute and manage input, weight, and output data, SRAM-CIM  accelerators typically follow a three-stage pipeline: 
(1) Buffering input data in Input SRAM, (2) Storing weights and computing in CIMs, and (3) Accumulating and buffering partial sums in Output SRAM.
Consequently, we constructed the Generalized Accelerator Template shown in Figure~\ref{Template} and extracted key parameters for compute-storage balancing.
The size of the input SRAM and output SRAM are denoted as (\textit{IS\_SIZE}) and (\textit{OS\_SIZE}) respectively. The CIMs array is organized in a grid format of macro rows (\textit{MR}) by macro columns (\textit{MC}). 
Outputs are accumulated along the row direction, while inputs are broadcast along the column direction. 
Besides, the accelerator has a data transfer bandwidth to external memory of Bandwidth (\textit{BW}) bits per cycle.

\subsection{Fine-grained Two-level Mapping Strategy}

% To address the mapping challenges of large-scale matrix multiplication operators on SRAM-CIM accelerators, we propose a fine-grained two-level mapping strategy comprising accelerator-level scheduling and macro-level tiling. 

\textbf{For accelerator-level scheduling,} we synthesize common strategies from previous CIM mapping works, encompassing both spatial and temporal approaches.
As shown in Figure~\ref{Scheduling} (a), for spatial scheduling, CIM-Tuner offers two options determining whether input activation or weight data should be stored as CIM weights.
We divide spatial scheduling into Non-Reversed (NR) and Reversed (R), where NR denotes storing the input activation in Input SRAM and R denotes storing the activation in CIM array. 
These correspond to the weight stationary and input stationary dataflow referenced in previous work~\cite{SXY-TCAD}. 
% Note that such spatial scheduling fundamentally determines whether a network's input or weight parameters reside in CIM macros. 
Meanwhile, previous temporal scheduling strategy typically follows the conventional input-priority update (IP) approach, as shown in Figure~\ref{Scheduling} (b), where input SRAM is updated first to maximize data reuse in CIMs.
CIM-Tuner introduces an alternative temporal scheduling strategy called weight-priority update (WP). WP prioritizes updating data in SRAM-CIM to enhance data reuse of input SRAM. 
% For temporal scheduling, we propose weight-priority (WP) and input-priority (IP) updating strategies.
% As the pseudo code in Algorithm 1 shows, R and NR strategies swap the storage locations of matrices A and B, while WP and IP strategies exchange the loop order outside CIM's computation.

\begin{figure}[!t]
\centering
\includegraphics[width=3.4in]{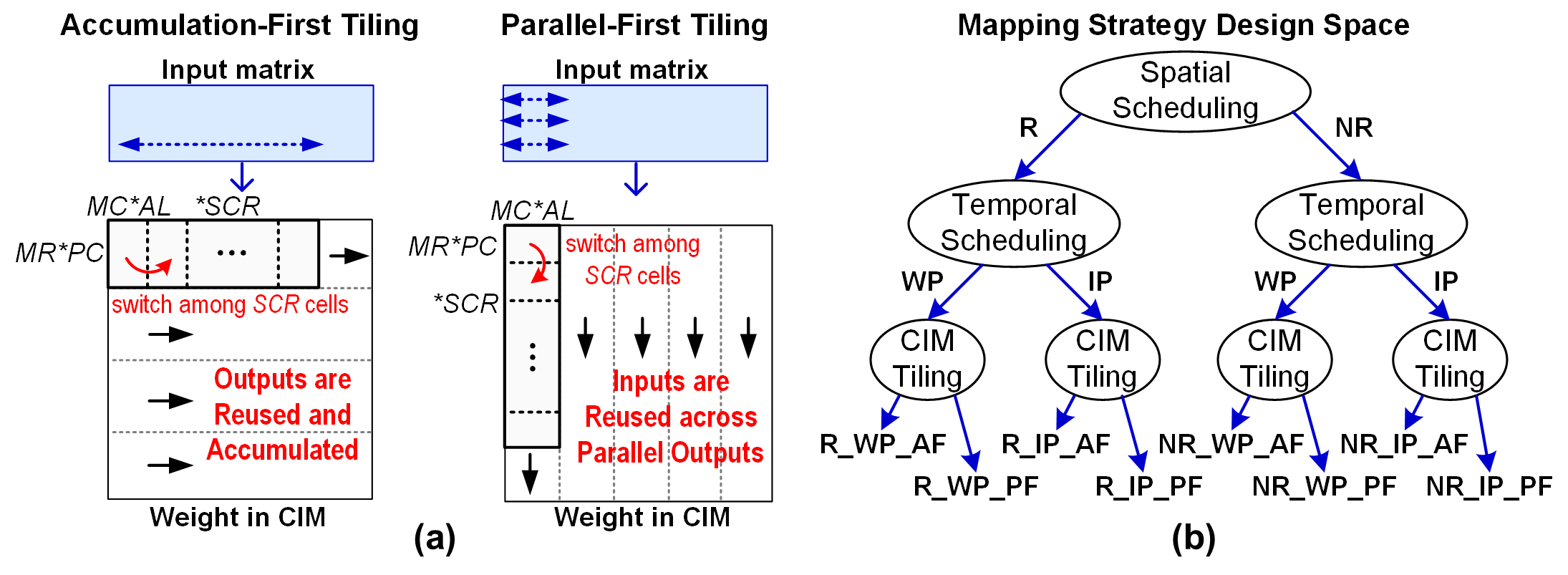}
\caption{(a) Accumulation-first (AF) tiling and parallel-first (PF) tiling during CIM computation.\ (b) Total mapping strategy design space. }\label{Mapping}
\end{figure}

\textbf{For macro-level tiling}, CIM-Tuner proposes two novel tiling strategies: accumulation-first tiling (AF) and parallel-first tiling (PF), in order to choose the optimal \textit{SCR} of CIM macro. 
As illustrated in Figure~\ref{Mapping} (a), The AF strategy prioritizes mapping \textit{SCR} CIM data blocks within the same output channel. This approach allows partial sums (Psum) generated in consecutive cycles to be reused and accumulated, though requiring distinct input vectors.
Conversely, the PF tiling prioritizes mapping \textit{SCR} blocks within the same input channel. This benefits from reusing input vectors across consecutive cycles but necessitates temporarily storing different partial sums in the output SRAM. These partial sums are later fetched and accumulated. 
The AF and PF strategies capture the inherent trade-off between CIM input and output, thereby expanding the overall mapping strategy space and unlocking further optimization potential.
The mapping strategy space comprising scheduling and tiling is shown in Figure~\ref{Mapping} (b). 
% Totally eight strategies can be explored when dealing with a matrix multiplication operator.

\begin{figure*}[!t]
\centering
\includegraphics[width=7in]{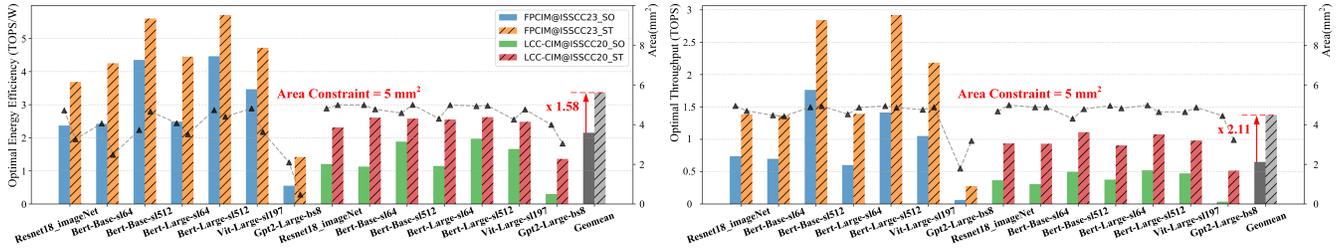}
\caption{Comparison with previous CIM mapping in~\cite{SXY-TCAD}. Optimal metrics are obtained through identical hardware-mapping co-exploration under same area budget of 5 mm$^2$. SO stands for spatial scheduling only~\cite{SXY-TCAD} and ST stands for both scheduling and tiling, i.e., mapping strategies in CIM-Tuner. }\label{Comparison}
\end{figure*}

\subsection{Optimization Objective and Hardware Space Pruning}
Since fine-grained mapping strategies are explored for each operator, the strategy space for the entire network grows exponentially with the number of operators. 
To address this challenge, our hardware-mapping co-exploration is designed to be operator-size-aware. This allows operators of the same size to be merged, significantly reducing the strategy space that the network must handle. 
% As a result, the total optimization objective for multiple operators can be expressed as a weighted sum of operators with different sizes.

% \begin{equation}
% F_{network} = \sum_i F_{size\_i} * count[size\_i]
% \end{equation}

When considering practical hardware implementation constraints, the continuous-valued hardware design space can also be pruned. First, to align with the address decoding, key hardware parameters including \textit{SCR}, \textit{IS\_SIZE}, and \textit{OS\_SIZE} must be constrained to powers of 2. Second, to ensure efficient utilization of the accelerator's external bandwidth, hardware designs where internal bandwidth (either \textit{ICW} or \textit{WUW}) falls below the specified \textit{BW} are eliminated.

% For exploring this optimized and pruned hardware-mapping co-design space, we employ the simulated annealing algorithm. The implementation details are available in our open-source repository. Section IV will present the experimental results demonstrating the effectiveness of CIM-Tuner.

\section{Experiment}
\subsection{Experiment Setup}
For the hardware template, we first implemented the proposed parameterized accelerator template using Verilog HDL. In this implementation, the CIM Macro is replaced with a parameterized behavior model. 
% The specific accelerator instruction set and source code can be found in the open-source repository. 
Next, the area and instruction power of several parameter configurations under the 28nm technology node can be obtained using DC Synthesize and Prime Time PX (PTPX) power analysis tools. 
Within the simulator, linear programming was applied to fit an instruction-level power model and the corresponding area model. 
To handle different CIM designs, we performed conversions in the simulator based on equivalent formulas from Matrix Abstraction. This approach ensures the adaptability 
% across various CIM configurations 
and normalizes all performance metrics to the same technique node. 

Regarding the mapping strategy, we developed the CIM-Tuner Compiler in C language, which compiles different mapping strategies into instruction flows based on the generalized accelerator template. 
A validation script is also provided to verify the functional correctness of these compiled instruction flows. 
For the co-exploration, we implemented the CIM-Tuner Simulator and simulated annealing algorithm in Python. 
The simulator analyzes instruction flows to assess the performance of the hardware accelerator across different operators, determining the optimal mapping strategy. 
Through the simulated annealing algorithm, hardware configurations are iteratively adjusted, enabling a co-exploration process that ultimately achieves an optimal hardware-mapping co-design.

\begin{figure}[!t]
\centering
\includegraphics[width=3.4in]{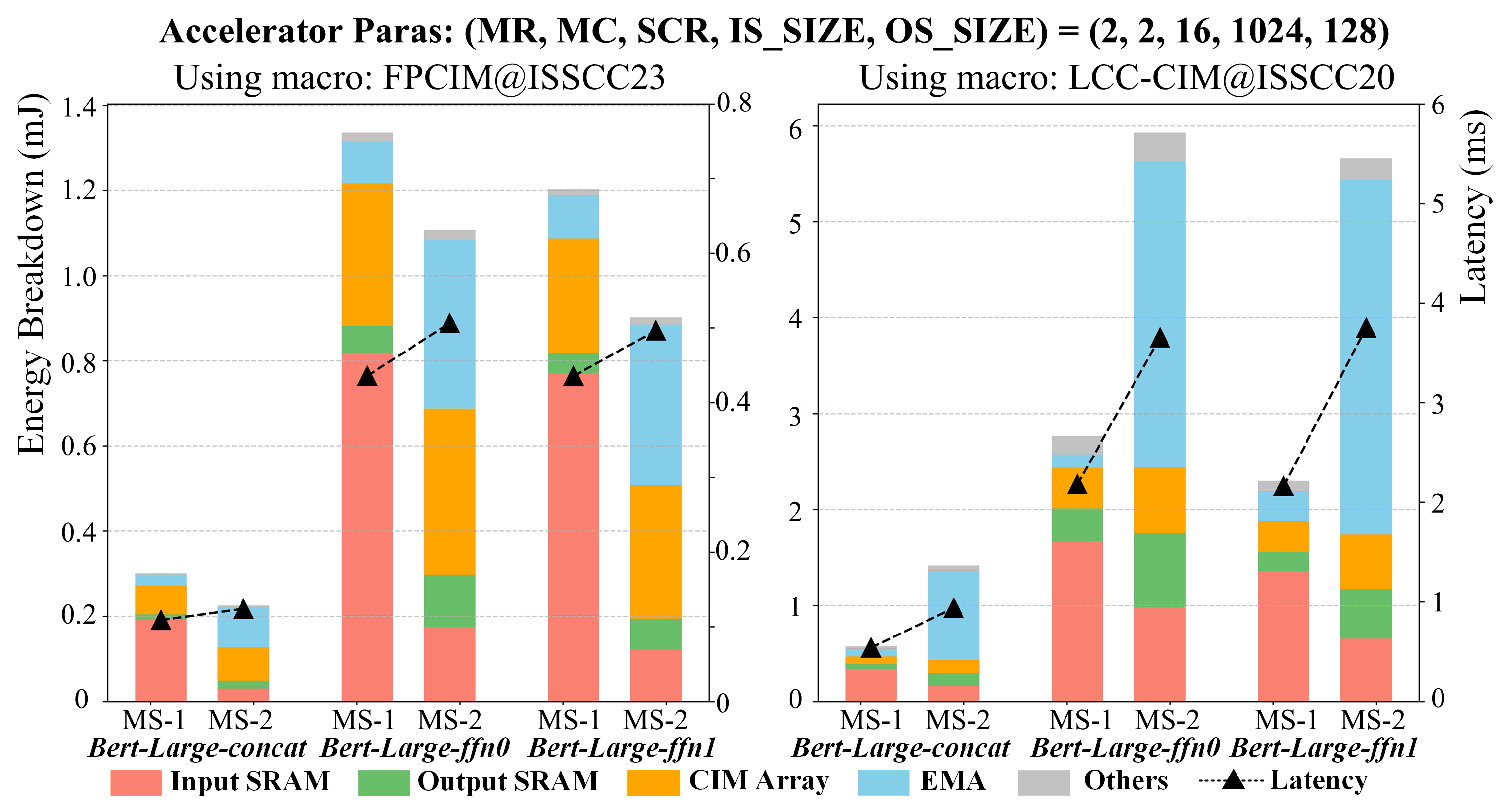}
\caption{Energy breakdown for Bert~\cite{Bert} operators under different CIM designs and Mapping Strategies. MS-1 stands for ``NR-IP-\textbf{AF}'' strategy and MS-2 stands for ``NR-IP-\textbf{PF}'' strategy. }\label{Breakdown}
\end{figure}

\subsection{Mapping Strategy Comparison}
To demonstrate the effectiveness of our mapping strategy, we selected SOTA CIM optimization work~\cite{SXY-TCAD} as the baseline. In~\cite{SXY-TCAD}, three optimizations for CIM accelerators are proposed:
(1) selecting weight/input stationary mapping strategies;
(2) adjusting macro dimensions; and
(3) modifying the macros floorplan.
Here, its mapping strategy corresponds to the spatial scheduling in CIM-Tuner. The latter two hardware modifications can be fully addressable through our hardware parameters exploration.
Thus, for reproducing~\cite{SXY-TCAD}, we utilized the same CIM-Tuner co-exploration framework but restricted the mapping strategy to spatial scheduling. As shown in Figure~\ref{Comparison}, across experiments on seven networks, CIM-Tuner demonstrated an average energy efficiency and throughput improvement of 1.58$\times$ and 2.11$\times$ compared to the baseline. 
% This exclusively proves that our extended mapping strategies provide additional optimization potential, enabling better hardware-mapping co-design. 

\begin{table*}[htbp]
\centering
\caption{CIM-Tuner Applied on SOTA Accelerators}
\begin{center}
\begin{tabular}{|c|c|c|c|c|c|}
\hline
Name & \textit{(MR, MC, SCR, IS\_SIZE, OS\_SIZE)} & Energy Eff.(TOPS/W) & Throughput(GOPS) & Area(mm$^2$) & Improve \\
\hline
TranCIM-Base & $(3, 1, 1, 64, 128)$ & $2.54$ & $1002.3$ & $3.52$ & / \\
\hline
TranCIM-EE. & $(2, 1, 16, 4, 64)$ & \textbf{$\textbf{3.40}$} & $734.3$ & $2.99$ & \textbf{$\times\textbf{1.34}$} \\
\hline
TranCIM-Th. & $(3, 1, 2, 2, 32)$ & $3.05$ & \textbf{$\textbf{1028.9}$} & $3.51$ & \textbf{$\times\textbf{1.03}$} \\
\hline
TP-DCIM-Base & $(2, 4, 1, 16, 16)$ & $1.89$ & $460.9$ & $2.23$ & / \\
\hline
TP-DCIM-EE. & $(3, 2, 4, 2, 2)$ & \textbf{$\textbf{4.36}$} & $1324.8$ & $2.04$ & \textbf{$\times\textbf{2.31}$} \\
\hline
TP-DCIM-Th. & $(3, 2, 4, 4, 256)$ & $4.25$ & \textbf{$\textbf{1326.7}$} & $2.16$ & \textbf{$\times\textbf{2.88}$} \\
\hline
\multicolumn{6}{l}{$^{\mathrm{*}}$EE. is for optimal energy efficiency and Th. is for optimal throughput. Other hardware parameters are fixed during exploration.}
\end{tabular}\label{tab1}
\end{center}
\end{table*}

Moreover, to analyze the gains of tiling strategy, a breakdown study is conducted on three matrix multiplication operators from the Bert-large~\cite{Bert} network across two CIM macros. 
The accelerator parameters \textit{(MR, MC, SCR, IS\_SIZE, OS\_SIZE)} are fixed at (2, 2, 16, 1024, 128). 
We selected "NR-IP-AF" and "NR-IP-PF" mapping strategies for comparison to demonstrate the impact of AF and PF strategies. 
As shown in Figure~\ref{Breakdown} and consistent with our earlier analysis, the AF strategy reduces Output SRAM overhead at the cost of increased Input SRAM overhead, whereas the PF strategy exhibits the opposite behavior. 
Given the limited 128 kB Output SRAM capacity, the PF strategy requires more external memory access (EMA) to store partial sums, significantly increasing the energy consumption. 
Compared to FPCIM, LCC-CIM's shorter accumulation length generates more partial sums for the same operator, resulting in more severe EMA penalties. 
% These findings confirm that different CIM tiling strategies yield substantial performance variations across hardware configurations and CIM types, validating the effectiveness of the tiling strategy proposed by CIM-Tuner.

\subsection{Improvement on SOTA accelerators}
To better and more directly demonstrate CIM-Tuner's ability to adjust the compute and storage capacity of the accelerator, we instantiated two state-of-the-art SRAM-CIM accelerators, TranCIM~\cite{TranCIM} and TP-DCIM\cite{TP-DCIM}. 
We extracted their corresponding CIM macro configurations and accelerator template parameters to serve as the baseline. We deliberately ignored their computational optimizations, focusing exclusively on the storage-compute ratio aspect of the accelerators.

\begin{figure}[!t]
\centering
\includegraphics[width=3.4in]{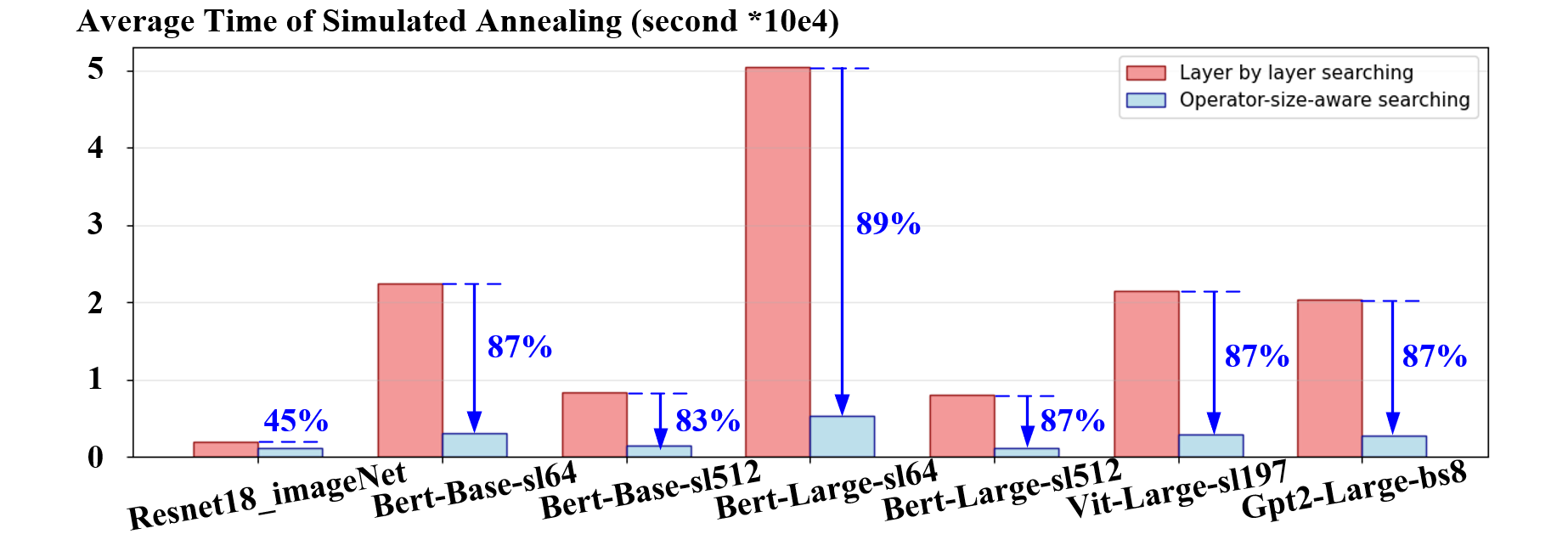}
\caption{Runtime illustration of simulated annealing.}\label{Runtime}
\end{figure}

As shown in Table~\ref{tab1}, we selected Bert-Large as the workload and performed co-exploration under different optimization targets. 
The area budgets are fixed as the baseline area. After co-exploration, the target computational capacity parameterized as \textit{MR and MC}, along with storage capacity parametrized as \textit{SCR, IS\_SIZE}, and \textit{OS\_SIZE}, are all adjusted. 
These adjustments demonstrated significant improvements. 
This outcome proves that the original accelerator design's storage-compute ratio allocation has room for enhancement, and the optimal mapping strategies for different operators across network should be modified accordingly.

We note that the performance metrics here show notable differences from the original accelerators' results. This divergence arises because the original metrics incorporated various computational optimizations and were tested under extreme conditions like low voltage and frequency. 
In contrast, CIM-Tuner's accelerator performance relies on our generalized template verification at standard test conditions. The power accuracy and functionality of our accelerator template are silicon-verified. Our approach prioritizes the performance gain through optimized compute and storage capacity balancing, while excluding specific computational optimizations.

\begin{figure}[!t]
\centering
\includegraphics[width=3.4in]{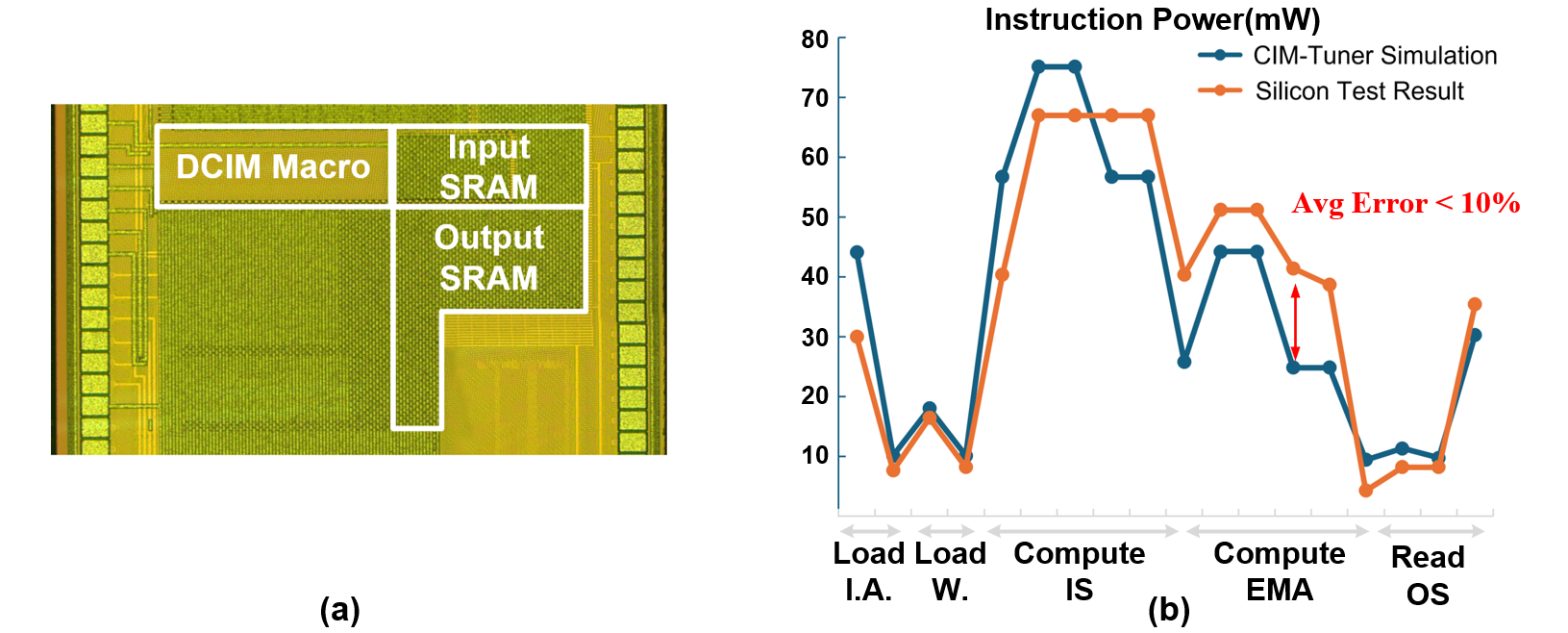}
\caption{Power accuracy verification on silicon results. (a) Die photo of instantiated SRAM-CIM accelerator. (b) Tested power of supported instructions. }\label{PAcc}
\end{figure}
\subsection{Runtime Overhead and Acceleration}
Figure~\ref{Runtime} shows the runtime of CIM-Tuner on seven workloads. The time is estimated as the average time of macro~\cite{FPCIM} and~\cite{LCC-6T} targeting energy efficiency and throughput under the area budget of 5 mm$^2$. By incorporating the operator-size-aware objective function, an average reduction of over 80\% in execution time is achieved. Furthermore, compared to the vanilla hardware search method in~\cite{SXY-TCAD}, our approach incorporating hardware pruning and bandwidth constraints reduces the total hardware design space by over 35\%.

\subsection{CIM-Tuner Accuracy}
% Since the CIM-Tuner's simulator is instruction-driven, the accuracy can be distinguished as cycle-level accurate as long as the instruction-level results is accurate. 
For the latency verification, a verification script is provided to examine the instruction flow of CIM-Tuner compiler. By analyzing the generated memory access address trace, the matrix multiplication flow can be directly verified. 
For the power and functionality, a 28nm prototype accelerator of \textit{(MR, MC, SCR, IS\_SIZE, OS\_SIZE)} $= (1, 1, 16, 16, 16)$ is silicon verified as the Figure~\ref{PAcc} shows. 
A vanilla DCIM macro is applied with parameters of \textit{(AL, PC, SCR, ICW, WUW)} $= (64, 8, 8, 512, 128)$. 
% The CIM macro is parametrized according to the matrix abstraction template. 
Compared to simulation results, a relative error of less than 10\% is observed, which is considered very accurate as the simulation and silicon difference.

\section{Conclusion}
The various CIM implementations and under-explored mapping strategy impede the SRAM-CIM accelerators hardware balancing exploration. 
CIM-Tuner tackles these challenges by introducing a generalized accelerator template and fine-grained two-level mapping strategies. Via hardware-mapping co-exploration, optimal performance can be achieved over SOTA CIM accelerators. 
% Open-sourced CIM-Tuner tool may help better early-stage accelerator decisions.

% Note that the purpose of this accelerator template is to investigate optimal compute-storage balancing. Consequently, the design targets the basic matrix multiplication. No specific optimizations are applied to the computing path, including special function units (SFU) for specific operators. This decision reflects our view that the processing of SFU represents a fixed computing overhead inherent to different dataflows, which is separate from the storage resource allocation problem.
% Critically, the hardware allocation decisions made by CIM-Tuner do not conflict with previous computing optimizations. These two aspects can be combined during actual accelerator design to achieve optimal end-to-end performance.

\section*{Acknowledgment}
This work was supported in part by the National Natural Science Foundation of China (62522403, 92264203, 62325405, U24B6015, 62504139); 
in part by the Key Research and Development Program of Jiangsu Province (BE2023020$-$1) and Beijing Natural Science Foundation (L257010); 
and in part by the Fundamental Research Funds for the Central Universities.

\end{document}